\documentstyle[jkas39]{article}

\runningauthor{KYEONG ET AL.} \runningtitle{STAR CLUSTERS OF THE DWARF IRREGULAR GALAXY IC 5152}
\beginpage{1}
\endpage{6}

\newcommand{\hi}{{\sc H~i}\/ }
\newcommand{\hii}{{\sc H~ii}\/ }

\def\arcdeg{\hbox{$^\circ$}}
\def\arcmin{\hbox{$^\prime$}}
\def\arcsec{\hbox{$^{\prime\prime}$}}

\begin{document}
\title{NEAR-INFRARED PHOTOMETRY OF THE STAR CLUSTERS IN THE DWARF IRREGULAR GALAXY IC 5152}

\author{
{\normalsize\textbf{\textsc{J}}\scriptsize\textbf{\textsc{AEMANN}}}
{\normalsize\textbf{\textsc{K}}}\scriptsize\textbf{\textsc{YEONG}}$^1$
\textbf{\textsc{,}}
{\normalsize\textbf{\textsc{E}}}\scriptsize\textbf{\textsc{ON}}$-${\normalsize\textbf{\textsc{C}}}\scriptsize\textbf{\textsc{HANG}}
{\normalsize\textbf{\textsc{S}}}\scriptsize\textbf{\textsc{UNG}}$^1$ 
\textbf{\textsc{,}}
{\normalsize\textbf{\textsc{S}}}\scriptsize\textbf{\textsc{ANG}}
{\normalsize\textbf{\textsc{C}}}\scriptsize\textbf{\textsc{HUL}}
{\normalsize\textbf{\textsc{K}}}\scriptsize\textbf{\textsc{IM}}$^1$
\textbf{\textsc{,}} \\
{\normalsize\textbf{\textsc{S}}}\scriptsize\textbf{\textsc{ANGMO}}
{\normalsize\textbf{\textsc{T}}}\scriptsize\textbf{\textsc{ONY}}
{\normalsize\textbf{\textsc{S}}}\scriptsize\textbf{\textsc{OHN}}$^{2,3}$
\textbf{\textsc{AND}}
{\normalsize\textbf{\textsc{H}}}\scriptsize\textbf{\textsc{YUN}}$-${\normalsize\textbf{\textsc{I}}}\scriptsize\textbf{\textsc{L}}
{\normalsize\textbf{\textsc{S}}}\scriptsize\textbf{\textsc{UNG}}$^1$
}

\offprints{J. Kyeong}
\address{$^1$Korea Astronomy \& Space Science Institute, Taejon 305-348, Korea}
\address{\it E-mail: jman,ecsung,sckim,hisung@kasi.re.kr}
\address{$^2$Center for Space Astrophysics, Yonsei University, Seoul 120-749, Korea}
\address{$^3$Space Astrophysics Lab, California Institute of Technology, MC 405-47, 1200 East California Boulevard, Pasadena, CA 91125}
\address{\it E-mail: tonysohn@srl.caltech.edu}

\vskip 3mm
\address{\normalsize{\it (Received Nov. 16, 2006; Accepted Dec. 1, 2006)}}

\abstract{
We present $JHK$-band near-infrared photometry of star clusters in the dwarf irregular 
  galaxy IC 5152. After excluding possible foreground stars, a number of candidate star 
  clusters are identified in the near-infrared images of IC 5152, which include young 
  populations. Especially, five young star clusters are identified in the $(J-H, H-K)$ 
  two color diagram and the total extinction values toward these clusters are estimated 
  to be $A_V =2 - 6$ from the comparison with the theoretical values given by the 
  Leitherer et al. (1999)'s theoretical star cluster model.}

\keywords{galaxies: individual (IC 5152) --- galaxies: dwarf irregular galaxies --- 
galaxies: star clusters --- galaxies: photometry - infrared: galaxies}

\maketitle

\section{INTRODUCTION}

Since the discovery of new Milky Way satellite galaxies 
  (e.g., Ursa Major dwarf spheroidal galaxy (Willman et al. 2005), 
  Canes Venatici dwarf galaxy (Zucker et al. 2006a),
  Bootes dwarf galaxy (Belokurov et al. 2006),
  Ursa Major II dwarf spheroidal galaxy (Zucker et al. 2006b)) and 
  new M31 satellite galaxies (e.g., 
  Andromeda IX dwarf spheroidal galaxy (Zucker et al. 2004),
  Andromeda X dwarf spheroidal galaxy (Zucker et al. 2006c)),
  the number of dwarf galaxies in the Local Group has been increased to
  well over 30 (Grebel 2000).
The dwarf galaxies are the most abundant type of galaxy in groups and clusters and 
   the building blocks of more massive galaxies. 
Furthermore the intrinsic properties of dwarf galaxies, 
  such as mass, density, and gas content, are likely to affect the star 
   formation history and chemical enrichment. 
Therefore, the understanding of the properties of dwarf galaxies
   provides clues on the galaxy formation processes.

IC 5152 (=IRAS 21594--51, ESO 237-27), a dwarf irregular galaxy (Sdm IV$-$V), 
  is one of the best objects available to 
  study starbursts due to its rather close distance.
On the other hand, a very bright star (HD 209142, $V=7.9$, $K_s=7.14$) 
   in the north-western part of the galaxy makes it difficult to obtain deep exposures. 
Basic information of IC 5152 is summarized in Table 1.

It has been controversial whether IC 5152 is a member of the Local Group or not. 
Sandage (1986) estimated a distance modulus of $(m-M)_0=26$ from the brightest stars 
  in this galaxy, and
van den Bergh (1994) pointed out that this distance estimate is too uncertain to find 
   the possibility of the Local Group membership. 
Zijlstra \& Minniti (1999) found the distance modulus of IC 5152 to be 
  ($m-M)_0=26.15 \pm 0.2$ using $VI$ color-magnitude diagram (CMD) 
  and red giant branch (RGB) tip method. 
But their photometry is not so deep to determine the RGB tip magnitude. 
Recently, using the HST snapshot survey of nearby galaxies, 
  Karachentsev et al. (2002) found the accurate magnitude of the tip of the RGB (TRGB) of 
  IC 5152 to be $I$(TRGB)$=22.58 \pm 0.16$ mag and 
  derived the distance to be $2.07 \pm 0.18$ Mpc, equivalent to $(m-M)_0=26.58 \pm 0.18$, 
  therefore, it is located at the outskirts of the Local Group. 

\begin{table*}[t]
\begin{center}
{\bf Table 1.}~~Basic Information of IC 5152\\
\vskip 3mm
{\small
\setlength{\tabcolsep}{1.2mm}
\begin{tabular}{llc} \hline\hline
  Parameter & Information & Reference \\
  \hline
$\alpha_{J2000.0}$, $\delta_{J2000.0}$ &
        22$^h$ 02$^m$ 41.$^s$5, $-51$\arcdeg~ $17'$ $47''$  & NASA/IPAC Extragalactic Database \\
$l, b$             & 343.\arcdeg92, $-50.\arcdeg19$         & NASA/IPAC Extragalactic Database \\
Type               & Sdm IV--V                              & Sandage \& Bedke 1985 \\
Foreground reddening, $E(B-V)$ & 0.00 mag                   & Zijlstra \& Minniti 1999 \\
Diameter           & 2.1 kpc                                & Huchtmeier \& Richter 1986 \\
Minor to major axis ratio of the disk, $(b/a)_{disk}$ & 0.6 & Sandage \& Bedke 1985 \\
Inclination, $i$                & 55\arcdeg                 & Huchtmeier \& Richter 1986 \\
Distance modulus, (m-M)$_0$    & $26.58 \pm 0.18$ mag       & Karachentsev et al. 2002 \\
Distance, $d$                  & $2.07 \pm 0.18$ Mpc (1$''$ = 10.0 pc) & Karachentsev et al. 2002 \\
Absolute total magnitude, $M_B$ & $-14.8$                   & Huchtmeier \& Richter 1986 \\
Radial velocity w.r.t. Local Group centroid, $V_{LG}$ & +53 km s$^{-1}$ & Huchtmeier \& Richter 1986 \\
Metallicity of the disk, $Z_{\rm disk}$     & 0.002         & Skillman, Kennicutt, \& Hodge 1989 \\
Total mass-to-light ratio, $(M_T/L_B)$  & 3.5               & Zijlstra \& Minniti 1999 \\
Number of candidate star clusters with $K \le 16.6$ & 20    & This study \\
Number of candidate young star clusters             &  5    & This study \\
\hline
\end{tabular}
} 
\end{center}
\end{table*}

The brightest \hii region \#A  has been known in IC 5152 and several other \hii regions 
  are known in the disk of this galaxy (Hidalgo-G{\'a}mez \& Olofsson 2002).  
IC 5152 is also an \hi rich dwarf (Buyle et al. 2006). 
The \hi image of the galaxy shows that \hi gas is spread out on the face of IC 5152. 
This suggested that there should be many newly formed young stellar populations on the disk. 
The young populations are also contained in the massive young clusters which are hardly 
   seen on optical images due to surrounding gas clouds.

Light at near-infrared wavelengths is less affected by dust absorption than 
   at visible wavelengths($A_K = 0.1 A_V$), making it easier to probe the heavily obscured 
   star forming regions and detect young clusters.
   
Since the data used in this study do not have enough angular resolution 
  to detect individual stars of IC 5152, 
the goal of this paper is to find star clusters and investigate their properties.
Especially, young star clusters with heavy reddening are investigated using the
   near-infrared wavelength characteristics.

This paper is composed as follows.
We describe our near-infrared observations and data reduction in Section II.
Section III present the results and analyses on star clusters in IC 5152 and
  a summary is given in Section IV.

\section{NEAR-INFRARED OBSERVATIONS AND DATA REDUCTION}

\begin{figure*}
\vskip 5mm
  \epsfxsize=8.9cm
  \epsfysize=8.9cm
  \centerline{\epsffile{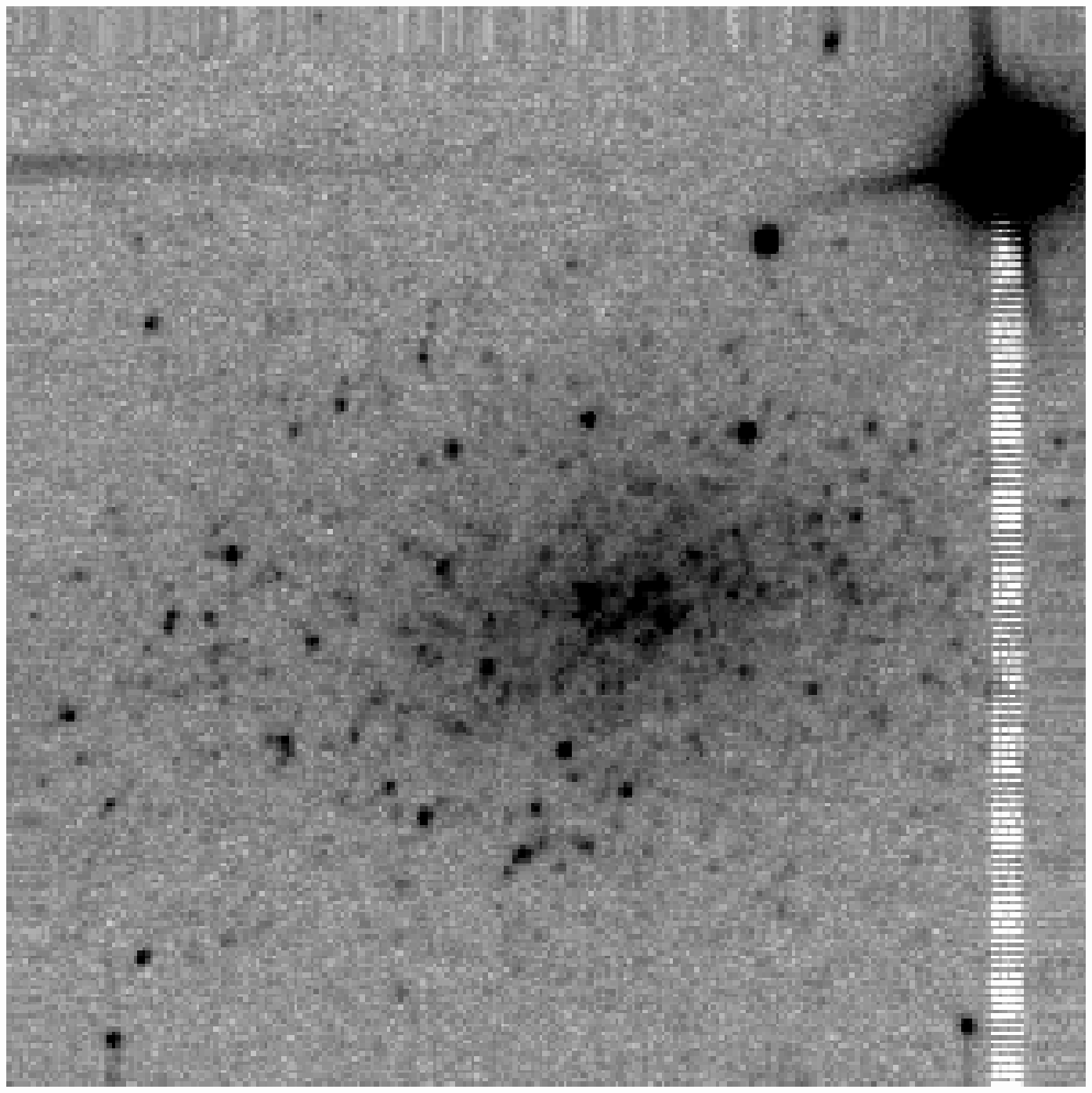}}
{\small {\bf ~~~Fig. 1.}---~Grey-scale map of the $K_n$-band CASPIR image of 
  IC 5152. 
North is at the top and east is to the left.
The size of the field is 2$\farcm$1 $\times$ 2$\farcm$1. 
The bright star on the north-western part of the image
  is HD 209142 ($V=7.9$, $K_s=7.14$).
}
\end{figure*}

$JHK_n$ images of IC 5152 were obtained on the night of UT 2002 June 30 
  using the 2.3 m telescope 
  and the infrared (IR) camera CASPIR (Cryogenic Array Spectrometer/Imager) 
  at the Siding Spring Observatory.  
A gray-scale map of the $K_n$-band CASPIR image of IC 5152 is displayed 
  in Figure 1.
The gain and readout noise of CASPIR are
  9 e$^-$/ADU and 50 electrons, respectively. 
CASPIR uses 256 $\times$ 256 InSb detector and the pixel scale is 
  $0\farcs50$ pixel$^{-1}$, which gives the field of view of
  2$\farcm$1 $\times$ 2$\farcm$1. 
In order to get high signal-to-noise ratio, a total of 9 frames with 5 s exposure 
  and 12 cycles were obtained, which 
  means that the combined image is a result of combining 108 frames of the same 
  exposure. 
The bias frames were frequently obtained over the night because the 
  bias level was known to vary throughout the night (McGregor 1995). 
Also, in order to remove possible detector instability and 
  temporal changes of the bright IR background, 
  nearby sky frames ($10\arcmin$ away from IC 5152) were taken with the same 
  exposure time. 

The instrument characteristics and the IR sky of strong signal and rapid variability,
  makes the reduction procedure of near-IR data more complex than that of optical CCD data. 
First, the raw data had to be linearized to recover the low and high intensity 
  information accurately according to the formula given by McGregor (1995). 
Then the bias and dark frames obtained just before and after each object frame 
  were subtracted. The median combined sky frame was subtracted from the object 
  frame.  
Finally, the data images are flattened by dome flats. 
Dome flats were 
  obtained for each filter by differencing exposures with the flatfield lamp on 
  and off, and several such frames were combined to form the final dome 
  flat for each filter.

We used the point spread function (PSF) fitting packages of DAOPHOT II and ALLSTAR for 
  the photometry.  
Stars of different frames were matched by DAOMATCH/DAOMASTER routines 
  (Stetson 1993).  
For each frame, several isolated unsaturated stars were used to construct 
  a good model PSF. 
Aperture corrections were made using the program DAOGROW (Stetson 1990) 
  for which we used the same stars used in the PSF construction.

In order to transform instrumental magnitudes to the standard system, we observed 12 SAAO
  photometric standard stars given by Carter \& Meadows (1995) 
  throughout the observing run.
We derived the transformation equations between our instrument magnitudes and the 
  standard values $J, H, K$ as following;
   
   \begin{eqnarray*}
     J &=& j + 18.87(\pm0.10) - 0.16(\pm0.09) * X_j -   \\
       & & -~~ 0.09(\pm0.07) * (J-K) \\
     H &=& h + 18.69(\pm0.03) - 0.14(\pm0.09) * X_h +    \\
       & & +~~ 0.20(\pm0.07) * (J-H) \\
     K &=& k + 17.79(\pm0.05) - 0.07(\pm0.13) * X_k +   \\
       & & +~~ 0.13(\pm0.09) * (J-K)
   \end{eqnarray*}
   \noindent
where $j$, $h$ and $k$ are instrumental magnitudes, $X_j$, $X_h$ and $X_k$ are airmasses
  at each bandpass.
The residuals of standard star calibration were 0.10, 0.03, and 0.06 mag for $J, H$, and
  $K$ filter, respectively. 
We can check the photometric calibration accuracy using a common star 
  observed by CASPIR (the second bright star in the $K$-band CASPIR frame 
  at X=175.8, Y=198.2) and 2MASS Point Source Catalog (Cutri et al. 2003). 
The comparison gives a good agreement in $J$ and $H$-bands within 
  the photometric errors. 
However, in the $K$-band, there is somewhat big difference of 
  $\Delta$(Ours$-$2MASS)=0.13 mag due to the different filter systems used, 
  $K_n$ and $K_s$.

\section{STAR CLUSTERS IN IC 5152}

  \subsection{Contamination of Foreground Stars}

\begin{figure*}
\centerline{\epsfxsize=14cm\epsfbox{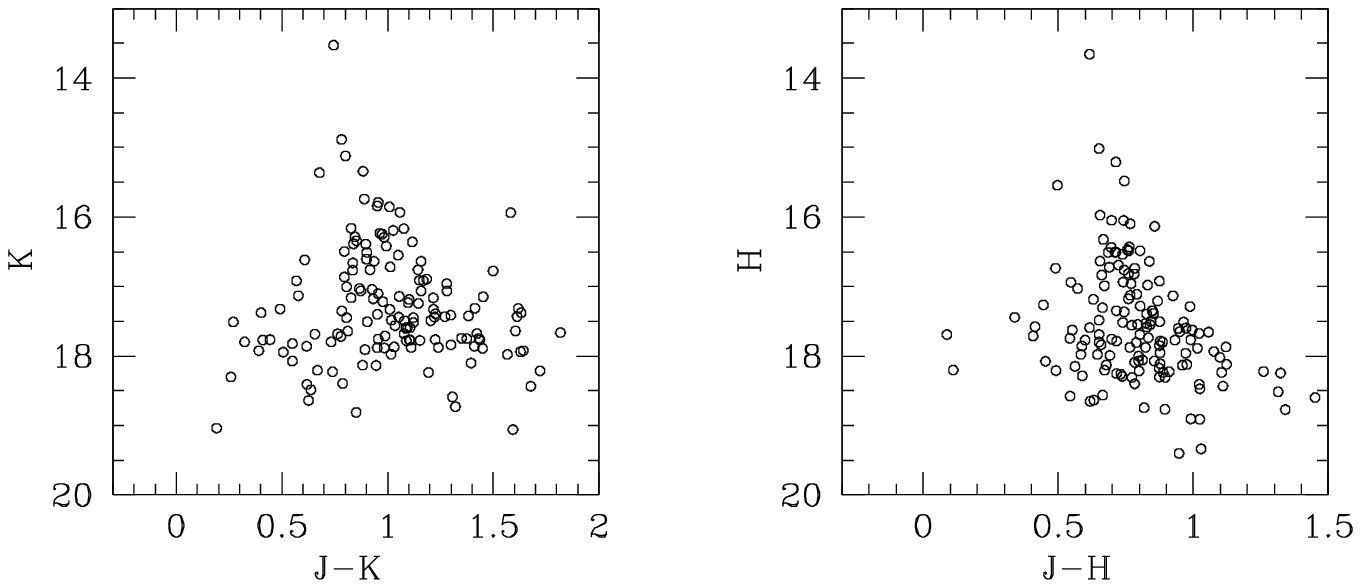}} \vskip 0.2cm \hskip 2.2cm
{\begin{minipage}{17.5cm} 
{\small {\bf ~~~Fig. 2.}---~~The color-magnitude diagrams of 
  all the objects detected in our near-infrared images of IC 5152. 
}
\end{minipage}}
\end{figure*}

We adopted the distance modulus of IC 5152 
  $(m-M)_0 =26.6 \pm 0.2$ (Karachentsev et al. 2002)
  and the reddening toward this galaxy $E(B-V) = 0.0$
  (Burstein \& Heiles 1984; Zijlstra \& Minniti 1999).
The color-magnitude diagrams for all the objects in IC 5152 
  are shown in Figure 2.

Since the detected objects are not resolved, they could be contaminated by foreground 
   stars.  
Likely foreground stars cannot be identified using near-IR CMD and/or 
   two-color diagram because their near-IR colors and brightnesses are not separated 
   from those of the IC 5152 members. 
Therefore, we identified the foreground stars using another catalog. 
As pointed out by Zijlstra \& Minniti (1999) already, 
  $2 \sim 3$ foreground stars are expected in our observed field of 
  2$\farcm$1 $\times$ 2$\farcm$1
  with $V\le 21$ mag from the Galactic star count model of Ratnatunga \& Bahcall (1985). 
One foreground star except the brightest star (HD 209142)
  at the upper right corner of the image can be identified using 2MASS Point Source Catalog.  
A comparison with sources in the 2MASS Point Source Catalog confirms the second 
  bright star($\alpha$(J2000) = 22$^h$ 02$^m$ 39.$^s$4, $\delta$(J2000) = 
  $-$51$^\circ$ 17$^\prime$ 05$\farcs$6, $J$ = 14.28, $H$ = 13.66, $K_s$ = 13.39).  

We also can identify the foreground stars in the blue plate (IIIaO + GG38) of NED 
  (NASA/IPAC Extragalactic Database)
  after subtracting out the diffuse body of the galaxy.  
A point-source like object is found during this process at X=28 and Y=29 of 
  our $K$-band image with $J$=16.89, $H$=16.09, and $K$=15.86 mag. 
Only very faint background galaxies might be included in our small field.

\subsection{Star Clusters}

   \begin{table}
     \begin{center}
      {\bf Table 2.}~~The candidated star clusters of IC 5152 ($K\leq$16.6) \\
         \begin{tabular}{lrrccc}
            \noalign{\smallskip}
            \hline \hline
            \noalign{\smallskip}
            ID   &X(pixels)  &Y(pixels) &$K$ & $J-H$ & $H-K$  \\
            \noalign{\smallskip}
            \hline
            \noalign{\smallskip}
            1  & 118.19 & 53.35  & 16.16  & 0.67  & 0.16  \\
            2  & 94.95  & 62.08  & 15.84  & 0.75  & 0.20  \\
            3  & 142.51 & 68.44  & 16.19  & 0.71  & 0.31  \\
            4  & 86.76  & 69.26  & 16.51  & 0.69  & 0.21  \\
            5  & 128.11 & 77.89  & 15.79  & 0.70  & 0.26  \\
            6  & 62.23  & 79.72  & 16.17  & 0.76  & 0.32  \\
            7  & 109.94 & 97.36  & 16.23  & 0.76  & 0.20  \\
            8  & 68.68  & 103.25 & 16.42  & 0.72  & 0.27  \\
            9  & 152.41 & 107.21 & 16.36  & 0.84  & 0.28  \\
           10  & 150.45 & 110.49 & 16.49  & 0.66  & 0.14  \\
           11  & 146.33 & 116.28 & 16.34  & 0.69  & 0.17  \\
           12  & 150.23 & 117.03 & 16.39  & 0.80  & 0.09  \\
           13  & 99.46  & 121.02 & 16.39  & 0.76  & 0.08  \\
           14  & 134.68 & 113.77 & 15.12  & 0.71  & 0.09  \\
           15$^\dagger$  & 49.62  & 123.92 & 15.34 & 0.75 & 0.14 \\
           16  & 101.86 & 149.01 & 16.29  & 0.74  & 0.29  \\
           17  & 171.16 & 152.84 & 14.88  & 0.65  & 0.13  \\
           18  & 133.78 & 156.07 & 15.74  & 0.66  & 0.23  \\
           19  & 75.44  & 159.58 & 16.55  & 0.78  & 0.27  \\
           20  & 30.76  & 178.70 & 16.29  & 0.70  & 0.15  \\
            \noalign{\smallskip}
            \hline
            \noalign{\smallskip}
          \end{tabular}

        \hspace{-4.5cm}$^{\dagger}$ : near the \hii region \#A
      \end{center}
   \end{table}

Using the magnitude of the brightest blue supergiants, $M_K = -10$ mag 
  (Rozanski \& Rowan-Robinson 1994) and our adopted distance modulus of IC 5152 
  $(m-M)_0 =26.6 \pm 0.2$ (Karachentsev et al. 2002), 
  the bright upper limit of supergiant magnitude is set to $m_K =16.6$ mag. 
The peak of the M31 globular cluster luminosity function exists near $M_K = -10$ 
  (Barmby et al. 2001), of which the globular cluster system is one of the
  best studied systems. We have detected 20 star cluster candidates brighter 
  than $K=16.6$ mag and listed them in Table 2 together with the $JHK$ photometry.

 In order to make direct confirmation for these star cluster candidates,
 we have examined the radial profiles of the candidate clusters with 
 the {\it HST/WFPC2} F814W image (Karachentsev et al. 2002). 
 Stars on the {\it WFPC2} WF3 image (pixel scale of 0.\arcsec1)
 have typically 0.9 pixels of FWHM. On the contrary, our sample of cluster candidates 
 have the FWHM range of 1.5 to 2.6 pixels. Thus, the result suggests that most of 
 the cluster candiates must be genuine star clusters. The young cluster candidates 
 found in the following subsection could not be confirmed due to the heavy 
 reddening in the optical wavelength.

\subsection{Young Clusters}

\begin{figure*}[t]
\centerline{\epsfxsize=14cm\epsfbox{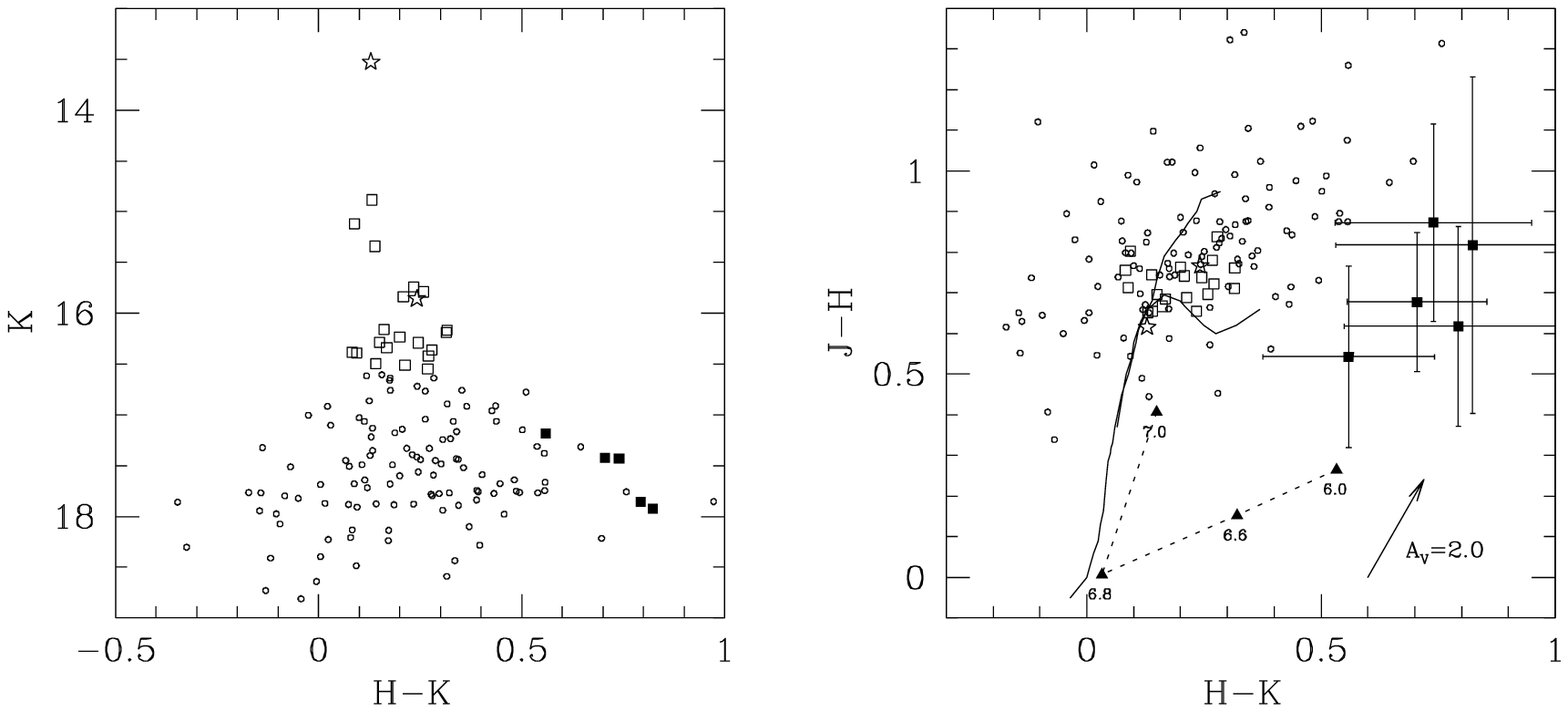}} \vskip 0.2cm \hskip 2.2cm
{\begin{minipage}{17.5cm} 
{\small {\bf ~~~Fig. 3.}---~~ ($K, H-K$) CMD({\it left}) and ($J-H, H-K$) 
  diagram({\it right}) of IC 5152. 
The symbols are ``star'' : field star, ``open squares'' : star clusters 
  and ``filled squares'' : young clusters.
The filled triangles mean the theoretical values given by Leitherer et al. (1999) 
  and the labels represent the assumed ages of log $t$ (see text for details). 
The error bars in the right panel show the estimated uncertainties 
  in the photometric calibration. 
The solid lines in the right panel show the sequences for solar neighborhood 
  giants (upper line) and dwarfs (lower line) given by Bessell \& Brett (1988). 
The arrow is the reddening vector $E(J-H) = 1.94 \times E(H-K)$ for
  $A_V = 2$ (Rieke \& Lebofsky 1985).
}
\end{minipage}}
\end{figure*}

It is plausible that there are many young star clusters in the star forming galaxy IC 5152.
However, it is not easy to identify young clusters only in CMD 
  because of the heavy internal reddenings.
Therefore, the $(J-H, H-K)$ two color diagram plotted in Figure 3 
  is very helpful for this purpose.
The suspected young clusters form a sequence in the ($J-H, H-K$) diagram that 
  parallels the reddening vector while the old clusters are located
  near the main sequence or giant branch lines given by Bessell \& Brett (1988). 
Five objects with very red colors are detected in Figure 3
  and these are identified as compact young clusters 
  forming a sequence that parallels the reddening vector.

   \begin{table}[t]
     \begin{center}
      {\bf Table 3.}~~The young cluster candidates and their reddening values in IC 5152 \\
       \begin{tabular}{rrcccr}
            \noalign{\smallskip}
            \hline \hline
            \noalign{\smallskip}
             X(pixels)  &Y(pixels) &$K$ & $J-H$ & $H-K$  & $A_V$ \\
            \noalign{\smallskip}
            \hline
            \noalign{\smallskip}
            40.68  & 29.08  & 17.86  & 0.62  & 0.79 & 2.5    \\
            159.26 & 105.20 & 17.92  & 0.82  & 0.83 & 4.5    \\
            139.98 & 115.61 & 17.19  & 0.54  & 0.56 & 3.0    \\
            60.41  & 119.80 & 17.43  & 0.87  & 0.74 & 5.6    \\
            129.98 & 192.77 & 17.42  & 0.68  & 0.71 & 3.7    \\
            \noalign{\smallskip}
            \hline
            \noalign{\smallskip}
         \end{tabular}
      \end{center}
   \end{table}

\begin{figure}[h]
\centerline{\epsfysize=8cm\epsfbox{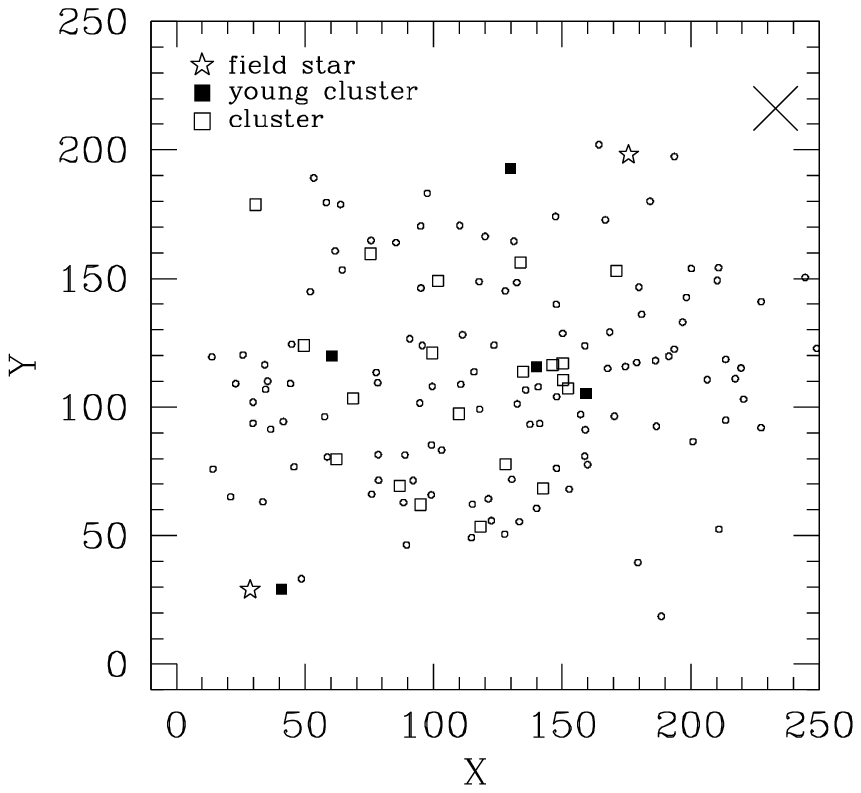}} \vskip 0.2cm \hskip 2.2cm
{\begin{minipage}{8.4cm} 
{\small {\bf ~~~Fig. 4.}---~~The identified objects in the observed field.
The symbols are same as in Figure 3, and the X and Y coordinates are 
  in pixel units.
The large cross in the upper right part represents the 
   brightest star HD 209142.
}
\end{minipage}}
\end{figure}

The line of sight extinction for each candidate young cluster was estimated by 
  extrapolation along the reddening vector to a point midway between the 
  log $t$(yr)=6.0 and 6.6 models given by Leitherer et al. (1999). 
The theoretical colors were computed by the $Z =0.004$ STARBURST99 models 
  with $\alpha =2.35$, $M_{\rm low}=1 M_\odot$, and $M_{\rm up} =100 M_\odot$, 
  where $Z$ is the metallicity, 
  $\alpha$ is the exponent of the power law initial mass function, and
  $M_{\rm low}$ and $M_{\rm up}$ are the low- and high-mass cutoff values,
  respectively.
While the metallicity derived for IC 5152 by Zijlstra \& Minniti (1999) is $Z=0.002$,
  we have adopted the metallicity of $Z =0.004$
  since the STARBURST99 models calculate only for five fixed values of metallicities.
    
The photometry result of the candidate young clusters and 
  their reddening values are listed in Table 3.
Especially, the young cluster near the \hii region \#A (X=60.41, Y=119.80) 
  shows heavy reddening ($A_V=5.6$). 
Finally the identified objects in the observed field are shown 
  in Figure 4.

\section{SUMMARY}

From the analysis of our $JHK$ photometry of IC 5152, we can summarize 
  the followings:
   
1. We found 20 star cluster candidates based on the distance modulus $(m-M)_0=26.6$
   and the bright limit of supergiant stars ($M_K = -10.0$) after excluding the 
   foreground stars that are expected from the Galactic star count model of 
   Ratnatunga \& Bahcall (1985). And the radial profiles of these candidates on 
   the {\it HST/WFPC2} image show that the FWHMs of these objects are much larger 
   than those of the typical stars. This confirms that these candidates must be 
   genuine star clusters. 
   
2. The possible young star clusters with heavy internal reddenings are identified 
   in the ($J-H, H-K$) two color diagram. Total extinction values toward these 
   young clusters are estimated to be $A_V =2 - 6$ from the comparison with 
   theoretical values.

\vspace{4mm}
We would like to thank the staffs of MSSSO, Australian National University 
   for the use of observing facilities. 
This research has made use of the NASA/IPAC Extragalactic Database (NED) 
  which is operated by the Jet Propulsion Laboratory, California Institute
  of Technology, under contract with the National Aeronautics and Space
  Administration.


\begin{thebibliography}{} 

\bibitem{} Barmby, P., Huchra, J. P., \& Brodie, J. P. 2001, 
  The M31 globular cluster luminosity function, AJ, 121, 1482
\bibitem{} Belokurov, V., et al. 2006, A faint new Milky Way satellite in Bootes,
  ApJ, 647, L111
\bibitem{} Bessell, M. S., \& Brett, J. M. 1988, 
  $JHKLM$ photometry -- Standard systems, passbands, and intrinsic colors, 
  PASP, 100, 1134
\bibitem{} Burstein, D. A., \& Heiles, C. 1984, 
  Reddening estimates for galaxies in the Second Reference Catalog and 
  the Uppsala General Catalog, ApJS, 54, 33
\bibitem{} Buyle, P., Michielsen, D., De Rijcke, S., Ott, J. \& Dejonghe, H. 2006,
  The CO content of the Local Group dwarf irregular galaxies IC 5152, UGCA 438, 
  and the Phoenix dwarf, MNRAS, 373, 793
\bibitem{} Carter, B. S., \& Meadows, V. S. 1995, 
  Fainter southern $JHK$ standards suitable for infrared arrays, MNRAS, 276, 734
\bibitem{} Cutri, R. M., et al. 2003, The IRSA 2MASS All-Sky Point Source Catalog, 
  http://irsa.ipac.caltech.edu/ \\ 
  applications/Gator/
\bibitem{} Grebel, E. K. 2000, The star formation history of the Local Group,
  in Star Formation from the Small to the Large Scale,
  Proceedings of the 33rd ESLAB symposium, eds. F. Favata, A. Kaas and A. Wilson, 
  (ESA, SP-445:ESA), 87
\bibitem{} Hidalgo-G{\'a}mez, A. M. \& Olofsson, K. 2002, 
  The chemical content of a sample of dwarf irregular galaxies, A\&A, 389, 836
\bibitem{} Huchtmeier, W., \& Richter, O. G. 1986, 
  \hi-observations of galaxies in the Kraan-Korteweg -- Tammann catalogue 
  of nearby galaxies. I. The data, A\&AS, 63, 323
\bibitem{} Karachentsev, I. D., Sharina, M. E., Makarov, D. I., Dolphin, A. E., 
  Grebel, E. K., Geisler, D., Guhathakurta, P., Hodge, P. W., Karachentseva, V. E., 
  Sarajedini, A., \& Seitzer, P. 2002, The very local Hubble flow,
  A\&A, 389, 812
\bibitem{} Leitherer, C., Schaerer, D., Goldader, J. D., G{\'o}nzalez Delgado, R. M., Robert, C., 
      Kune, D. F., de Mello, D. F., Devost, D., \& Heckman, T. M.
  1999, Starburst99: Synthesis Models for Galaxies with Active Star Formation,
  ApJS, 123, 3
\bibitem{} McGregor, P. 1995, Users Manual for the CASPIR on the MSSSO 2.3 m Telescope
\bibitem{} Ratnatunga, K. U., \& Bahcall, J. N. 1985, 
  Estimated number of field stars toward Galactic globular clusters and 
  Local Group Galaxies, ApJS, 59, 63
\bibitem{} Rieke, G. H., \& Lebofsky, M. J. 1985,
  The interstellar extinction law from 1 to 13 microns, ApJ, 288, 618
\bibitem{} Rozanski, R., \& Rowan-Robinson, M. 1994, 
  The accuracy of the brightest stars in galaxies as distance indicators,
  MNRAS, 271, 530
\bibitem{} Sandage, A. 1986, The redshift-distance relation. IX. Perturbation of 
  the very nearby velocity field by the mass of the Local Group, ApJ, 307, 1
\bibitem{} Sandage, A., \& Bedke, J. 1985,
   Candidate galaxies for study of the local velocity field and distance scale 
     using Space Telescope. I. The most easily resolved, 
  AJ, 90, 1992
\bibitem{} Skillman, E. D., Kennicutt, R. C., \& Hodge, P. W. 1989,
  Oxygen abundances in nearby dwarf irregular galaxies, ApJ, 347, 875
\bibitem{} Stetson, P. B. 1990, 
  On the growth-curve method for calibrating stellar photometry with CCDs, 
  PASP, 102, 932
\bibitem{} Stetson, P. B. 1993, Further progress in CCD photometry, in 
  Stellar Photometry, Current Techniques and Future Developments, 
  Proceedings of the IAU Colloquium No. 136, 
  eds. C. J. Butler \& I. Elliot(Cambridge University Press: Cambridge), 291
\bibitem{} van den Bergh, S. 1994, The outer fringes of the Local Group, AJ, 107, 1328
\bibitem{} Willman, B., et al. 2005, ApJ, A new Milky Way dwarf galaxy in Ursa Major,
  ApJ, 626, L85
\bibitem{} Zijlstra, A. A., \& Minniti, D. 1999, A dwarf irregular galaxy at the edge 
  of the Local Group: Stellar populations and distance of IC 5152, AJ, 117, 1743
\bibitem{} Zucker, D. B., et al. 2004, Andromeda IX: a new dwarf spheroidal satellite
  of M31, ApJ, 612, L121
\bibitem{} Zucker, D. B., et al. 2006a, A new Milky Way dwarf satellite in Canes Venatici,
  ApJ, 643, L103
\bibitem{} Zucker, D. B., et al. 2006b, A curious Milky Way satellite in Ursa Major,
  ApJ, 650, L41
\bibitem{} Zucker, D. B., et al. 2006c, Andromeda X, a new dwarf spheroidal galaxy of 
  M31: Photometry, ApJL, submitted (astro-ph/0601599)

\end{thebibliography}
\end{document}